\documentclass[twocolumn]{aastex631}

\usepackage{graphicx}	
\usepackage{amsmath}	

\usepackage{subfigure}

\shorttitle{Hidden binaries in star-forming regions}
\shortauthors{M. H. Rawcliffe \emph{et al.}}

\begin{document}

\title{Hidden binaries in star-forming regions}

\correspondingauthor{Richard Parker}
\email{R.Parker@sheffield.ac.uk}

\author{Mary H. Rawcliffe}
\affiliation{Astrophysics Research Cluster, School of Mathematical and Physical Sciences, University of Sheffield, Hicks Building, Sheffield S3 7RH, UK}

\author{Nathan Griffiths-Janvier}
\affiliation{Astrophysics Research Cluster, School of Mathematical and Physical Sciences, University of Sheffield, Hicks Building, Sheffield S3 7RH, UK}

\author[0000-0002-1474-7848]{Richard J. Parker}
\altaffiliation{Royal Society Dorothy Hodgkin Fellow}
\affiliation{Astrophysics Research Cluster, School of Mathematical and Physical Sciences, University of Sheffield, Hicks Building, Sheffield S3 7RH, UK}

\begin{abstract}
  A significant fraction of, and possibly all, stars form in binary or multiple systems. For Solar-mass stars in the Galactic field, the distribution of orbital separations is log-normal over seven orders of magnitude, from $10^{-2} - 10^5$\,au. In contrast, the separation distributions of systems in nearby star-forming regions paints a much more confusing picture. There appears to be an excess of systems in the separation range 10 -- 1000\,au, and recent high-resolution spectroscopic observations of close ($<10$\,au) systems suggest a field-like distribution in some star-forming regions, but a possible excess with respect to the field in other regions. Furthermore, the resolution limit of numerical simulations of binary star formation is $\sim$1\,au, and  consequently comparisons with the binary distributions in star-forming regions and in the field are restricted. In this paper, we demonstrate that these observational uncertainties, and limitations in the simulations, are potentially a much bigger problem than previously realised. We show that the log-normal separation distribution in the field can be reproduced by combining constituent binary populations whose initial separation distributions have a very different form to a log-normal. We also argue that the observed excess of binaries in the range 10 -- 62\,au in the ONC compared to the Galactic field is not necessarily inconsistent with the field population, because the ONC is only one of many star-forming regions that populate the field. We propose that further observations of spectroscopic binaries in star-forming regions to probe and complete the $<10$\,au parameter space are urgently needed. 
\end{abstract}



\section{Introduction}

A significant proportion of stellar systems in the Galactic field are found in binary or multiple systems \citep[30 -- 90\,per cent, depending on the primary mass in the system,][]{Duquennoy91,Raghavan10,Sana13,DeRosa14,Ward-Duong15,Tokovinin08,Tokovinin14}, and the fraction of binary and multiple systems in star-forming regions is higher still \citep{Reipurth07,Chen13}.

However, the statistics for binary and multiple systems in star-forming regions are highly incomplete, especially when compared to similar properties in the Galactic field \citep{King12a,King12b}. For example, the range of orbital separations for Solar-mass primary systems in the field spans seven orders of magnitude, from $10^{-2} - 10^5$\,au. In contrast, in the Taurus star-forming region the observed range is between $20 - 2000$\,au \citep{Kohler98}, and in the more distant (and crowded) ONC the observed range is between $10 - 620$\,au \citep{Reipurth07,Duchene18}.

Recent work by \citet{Kounkel19} analysing spectroscopic binaries in the APOGEE-2 data in a  handful of star-forming regions has suggested that close binaries in these nearby star-forming regions may follow a similar period distribution to the field. However,  observations of close binaries in young moving groups find an excess of spectroscopic binaries compared to the field \citep{Zuniga21}. 

Complicating matters further is the generally accepted notion that dynamical encounters in star-forming regions (which are orders of magnitude more dense than the Galactic field) change the orbital separation distribution of binary systems \citep[e.g.][and many others]{Kroupa95a,Kroupa95b,Kroupa99,Parker11c,Marks12,Parker23d,Cloutier24}, as well as destroying the wider systems and thereby reducing the  overall fraction of binary or multiple systems.

Furthermore, the uncertainty over whether spectroscopic binaries in star-forming regions should match the distribution in the field is exacerbated by the resolution  limit of hydrodynamical simulations that follow the formation of stellar binaries. These simulations do not yet make predictions for the abundance of very close binary systems as no binaries with separations less than $\sim 1$\,au can form in these simulations \citep[e.g.][]{Bate12,Bate14}.

Given this lack of observational consensus, and absence of simulation data for close stellar binaries, it is usually assumed that the orbital separation distributions in star-forming regions follow the log-normal distribution of binary separations in the Galactic field \citep{Raghavan10}. However this is the distribution for binaries with a Solar-mass primary star, and the peak of this distribution (i.e. the mode of binary separations) depends on primary mass \citep{Ward-Duong15,DeRosa14,Sana13}.

Because close ($<10$\,au) binary systems are dynamically `hard' \citep{Heggie75,Hills75a,Hills75b}, they are not susceptible to destruction and it is often assumed that the separation distribution of close binaries must match the distribution of close binaries in the Galactic field \citep{Kroupa95a,Kroupa95b}. Furthermore, it is thought that the close binary distribution provides direct information on the outcome of the star formation process, as these systems are unaffected by dynamical encounters. 

However, the binary population in the field must be the sum of binary distributions from many different star-forming regions, and so the individual binary properties (including the numbers of close binaries) may vary from region to region, and may produce (slightly) different distributions of close separation binaries. In addition, several theoretical works \citep{Stahler10,Korntreff12}  have argued that the orbital distribution of $<10$\,au binaries is actually very different to that in the Galactic field, with a significant excess of these systems immediately after formation, which then decay due to gas friction and merge to form single stars (and hence would not be included in the observed binary separation distribution at later ages).

In this paper we propose that the close binary population in star-forming regions is so uncertain that it should be targetted by observations as a matter of urgency. In Section~\ref{sec:constraints} we review the current observational and theoretical work on binary separation distributions in star-forming regions. In Section~\ref{sec:analysis} we demonstrate how different separation distributions in individual star-forming regions could be in comparison to the Galactic field population, we discuss our results in Section~\ref{sec:discussion} and we present conclusions in Section~\ref{sec:conclusions}.

\section{Observational and theoretical constraints}
\label{sec:constraints}

The primary detection method for short period (close) binary systems is via the radial velocity technique, and this has only been used sparingly to detect binaries in star-forming regions. Instead, the main detection method in star-forming regions is via visual imaging of companions. Although dependent on the distance to the region in question, observations typically probe the separation range 10 -- 1000\,au, and it is unclear whether the unseen binary populations in the regions thus far observed are consistent with the Galactic field population.

The Heggie-Hills law \citep{Heggie75,Hills75a,Hills75b} compares the binding energy of an individual binary to the typical encounter energy in a star-forming region to determine how likely it is that the binary in question will be destroyed. Close (``hard'', or ``fast'') binary systems are unlikely to be destroyed in star-forming regions (though their orbital separations may decrease), whereas wide (``soft'' or ``slow'') systems are highly likely to be destroyed. The hard-soft boundary depends on the stellar density and velocity dispersion in a star-forming region, but for most star-forming regions the observed visual binaries straddle the hard-soft boundary (typically tens to hundreds of au), and for these systems it is unclear whether we are oberving the outcome of star-formation, or whether these systems have undergone significant alteration due to dynamical encounters.

In the following, we compare the separation distribution of visual binaries in the ONC and Taurus to the separation distributions of binaries in the local Solar neighbourhood. The histograms of the separations of field binaries are typically fit with a Gaussian or log-normal \citep[see e.g.][though there is some deviation from a Gaussian in the separation distribution in most of these studies]{Duquennoy91,Fischer92,Raghavan10,Bergfors10,Janson12,Ward-Duong15,Cifuentes25}, and are fit with two parameters, the (log) mean separation, ${\rm log}_{10}\,\bar{a}$ and the variance,  $\sigma_{{\rm log}_{10}\,\bar{a}}$. These log-normal fits are then normalised to the overall multiplicity fraction $f_{\rm mult}$  of the sample in question, given by
\begin{equation}
f_{\rm mult} = \frac{B + T + ...}{S + B + T + ...},
\end{equation}
where $S$, $B$ and $T$ are the numbers of single, binary and triple systems.

In Table~\ref{field_props} we show ${\rm log}_{10}\,\bar{a}$, $\sigma_{{\rm log}_{10}\bar{a}}$ and $f_{\rm mult}$  from several works. The overall consensus is that the multiplicity fraction, and the peak, or mean, separation decreases as a function of the primary mass component in a binary system \citep{Duchene13b}, although again there is some uncertainty due to limitations in the separation ranges probed by the observations. For example, \citet{Ward-Duong15} see an increasing number of binaries with short separations with primary mass, and their `unrestricted' fit to the data implies a peak at much smaller separations ($\sim0.2$\,au) than previous work \citep[$\sim16$\,au,][]{Bergfors10,Janson12}, which itself is a revision downward from the first M-dwarf binary survey paper \citep[which found the peak of the separation distribution to be at 30\,au,][]{Fischer92}.

\begin{table*}
  \caption{Literature log-normal fits to the binary populations observed in the Galactic field. We show the spectral type of the primary mass, the main sequence mass range this corresponds to, the multiplicity fraction $f_{\rm mult}$, the mean separation $\bar{a}$, and the mean (${\rm log}_{10}\,\bar{a}$) and variance ($\sigma_{{\rm log}_{10}\,\bar{a}}$) of the log-normal fits to these distributions.}
\begin{center}
\begin{tabular}{cccccccc}
\hline 
Type & Primary mass & $f_{\rm mult}$ & $\bar{a}$ & ${\rm log}_{10}\,\bar{a}$ & $\sigma_{{\rm log}_{10}\,\bar{a}}$ & Ref. & Line in Figs.~\ref{observations}~and~\ref{theory_simulations} \\
\hline
M- & $0.08 < m_p/$M$_\odot \leq 0.45$ & 0.42 & 30\,au & 1.48 & 1.53 & \citet{Fischer92} & N/A \\
\hline
M- & $0.08 < m_p/$M$_\odot \leq 0.45$ & 0.34 & 16\,au & 1.20 & 0.80 & \citet{Bergfors10,Janson12} & Dashed blue \\
\hline
M- & $0.08 < m_p/$M$_\odot \leq 0.45$ & 0.65 & 0.22\,au & -0.66 & 1.86 & \citet{Ward-Duong15} & Solid blue \\
\hline 
G- & $0.8 < m_p/$M$_\odot \leq 1.2$ & 0.58 & 30\,au & 1.48 & 1.53 & \citet{Duquennoy91} & Dashed red \\
\hline
G- & $0.8 < m_p/$M$_\odot \leq 1.2$ & 0.46 & 50\,au & 1.70 & 1.68 & \citet{Raghavan10} & Solid red \\
\hline
\end{tabular}
\end{center}
\label{field_props}
\end{table*}

\begin{figure*}
   \subfigure[]{\includegraphics[scale=0.55]{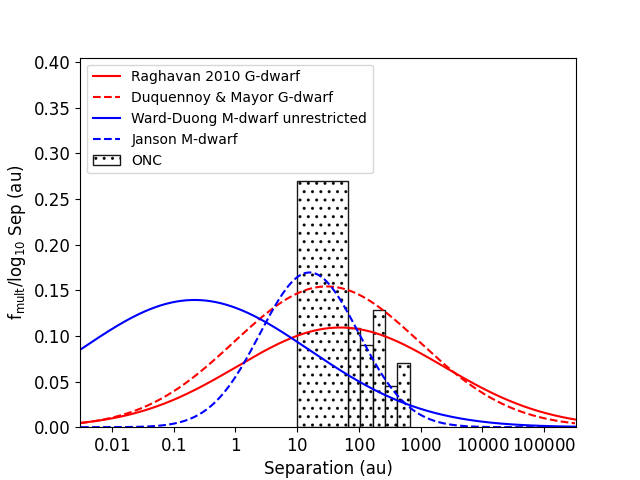}}
   \subfigure[]{\includegraphics[scale=0.55]{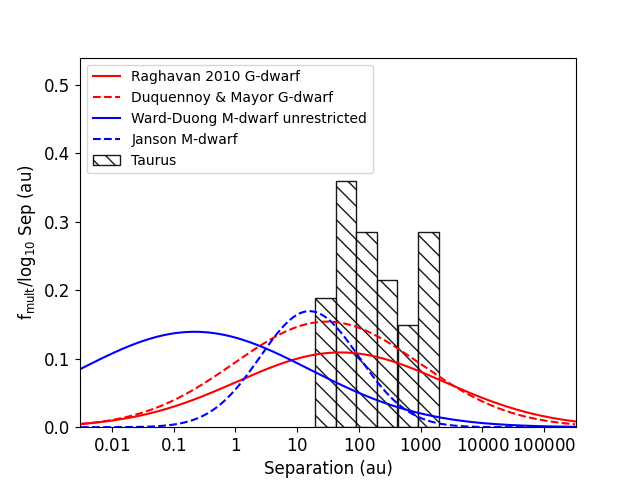}}
        \caption{Observations of binary separations in two star-forming regions. Data for visual binaries in the ONC are shown in panel (a); the leftmost bin is from \citet{Duchene18} and the other bins are from \citet{Reipurth07}. Data for visual binaries in Taurus \citep{Kohler98} are shown in panel (b). In both panels we also show two different fits to the observed G-dwarf Galactic field population from \citet[][the solid red line]{Raghavan10} and \citet[][the dashed red line]{Duquennoy91}, and two different fits to the observed M-dwarf Galactic field population from \citet[][the solid blue line]{Ward-Duong15} and \citet[][the dashed blue line]{Janson12}.}
     \label{observations}
 \end{figure*}

In Fig.~\ref{observations}, we show the separation distributions for binaries in the ONC (panel a) and Taurus (panel b). The most recent observations of the ONC \citep{Duchene18} appear to show a significant excess of binaries in the range 10 -- 62\,au, compared to the Galactic field. Taurus also has an excess of binaries compared to the Galactic field \citep{Kohler98}, but in the full separation range probed by the observations (20 -- 2000\,au). For comparison, the log-normal fits to the G-dwarf field binary separation distributions are shown by the solid red \citep{Raghavan10} and dashed red \citep{Duquennoy91} lines, and the log-normal fit to the field M-dwarf binary separation distribution \citep{Bergfors10,Janson12} is shown by the blue dashed line. A fit to the M-dwarf field binaries taking into account possible observational incompleteness \cite{Ward-Duong15} is shown by the solid blue line. 

 The excess of wider binaries in Taurus has been explained as being due to dynamical destruction of binary and multiple systems in their natal star-forming regions. This led to the hypothesis of a `universal' binary separation distribution, proposed to be the outcome of the star formation process and then dynamically altered to varying degrees, depending on the density of the star-forming region in question \citep{Kroupa95a,Kroupa95b,Kroupa99,Kroupa11,Marks11,Marks12}. \citet{Kroupa95b} argues that most star-forming regions are therefore dense, because the field population (likely to be the sum of many different star-forming regions) does not contain an excess of wide systems, and in this hypothesis Taurus is an outlier due to its excess of wide systems compared to the field.

\begin{figure}
 	\includegraphics[width=\columnwidth]{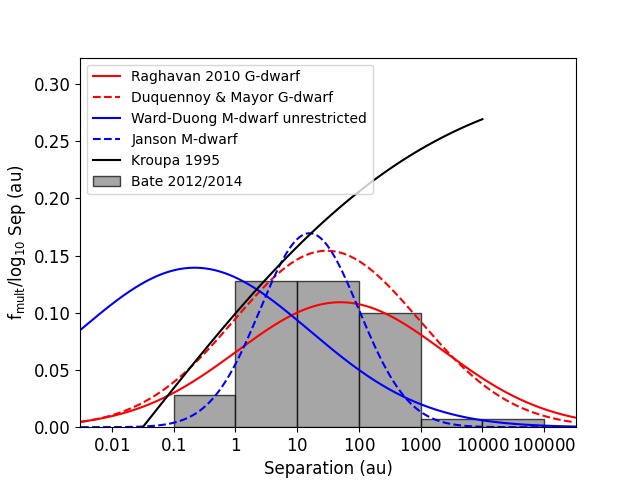}
        \caption{The distribution of binary separations from hydrodynamical simulations of star formation. The simulations shown are from \citet[][the solid histogram]{Bate14}, but results from an earlier simulation presented in \citet{Bate12} are very similar. For comparison, we show the proposed initial `universal' binary distribution from \citet{Kroupa95a} inferred from $N$-body simulations (the solid black line), as well as two different fits to the observed G-dwarf Galactic field population from \citet[][the solid red line]{Raghavan10} and \citet[][the dashed red line]{Duquennoy91}, and two different fits to the observed M-dwarf Galactic field population from \citet[][the solid blue line]{Ward-Duong15} and \citet[][the dashed blue line]{Janson12}.}
     \label{theory_simulations}
 \end{figure}

 The `universal' initial binary separation distribution (which also assumes all stars form in multiples, i.e.\,\,$f_{\rm mult} = 1$) is shown by the solid black line in Fig.~\ref{theory_simulations}. (Note that  this distribution was orginally formulated in terms of the \emph{period} distribution -- \citet{Parker14d} and \citet{Ballantyne21} recast this in terms of the separation, thus:
 \begin{equation}
f\left({\rm log_{10}}a\right) = \eta\frac{{\rm log_{10}} a - {\rm log_{10}} a_{\rm min}}{\delta + \left({\rm log_{10}} a - {\rm log_{10}} a_{\rm min}\right)^2},
\label{coma}
\end{equation}
where ${\rm log_{10}} a$ is the logarithm of the semi-major axis in au and ${\rm log_{10}} a_{\rm min} = -2$ [$a_{\rm min} = 0.01$\,au]. The numerical constants are $\eta = 5.25$ and $\delta = 77$.) 

 The `universal' initial binary distribution was inferred from comparing the evolution of binary populations in $N$-body simulations of star clusters. Because binaries with close ($<10$\,au) separations are dynamically `hard' they are unlikely to be destroyed, and so the separation distribution for close binaries was assumed to match the \citet{Duquennoy91} distribution.

 Other theoretical constraints on the binary separation distribution come from hydrodynamical simulations. In Fig.~\ref{theory_simulations} we also show the separations from the radiation hydrodynamical simulations of star cluster formation presented in \citet{Bate12} and \citet{Bate14}. These simulations have a resolution limit at around 0.5\,au, due to the size of the sink particles used to reduce the computational expense of the calculation. Whilst far from conclusive, these simulations seem to hint at an excess of binaries at small separations compared to the Galactic field G-dwarf systems (compare the histogram to the solid red line in Fig.~\ref{theory_simulations}.


 Finally, \citet{Stahler10} and \citet{Korntreff12} also argue for an excess of binaries with small separations immediately after star formation, on the basis that some of these decay and merge due to dynamical friction with the gas in their natal pre-/proto-stellar core.

\section{The Field binary population as the sum of star-forming regions}
\label{sec:analysis}

The log-normal distribution of binary separations in the field is sometimes characterised as being the sum of multiple different distributions overlaid on top of each other. It is usually assumed that these distributions would have a similar shape to the underlying field distribution. In the following we show that it is relatively straightforward to reproduce the symmetrical log-normal distribution in the field by combining several asymmetric (e.g.\,\,Maxwell-Boltzmann) distributions with different peak semimajor axes.

The distribution of wider binary orbits ($>10$\,au) in the field has been fit by functions similar to Maxwell-Boltzmann distributions \citep{ElBadry21,Makarov25}, and although some theoretical arguments rule out the binary separation distribution as being a true Maxwell-Boltzmann distribution (see the discussion in \cite{Makarov25} of the separation distributions in \cite{Jeans19} and \cite{Ambartsumian37}), the mathematical form of its probability distribution is a convenient way to produce peaked distributions with a tail that are similar to wide systems ($>10^3$\,au) observed in the field, and the excess of intermediate separation binaries (1 -- 100\,au) formed in numerical simulations \citep{Bate12,Bate14}.

To demonstrate that the sum of different types of one distribution shape can lead to an overall distribution with a different shape, we draw five sets of $N = 2000$ binaries, each with separations drawn from a Maxwell-Boltzmann--like distribution with a probability density distribution of the form: 
\begin{equation}
  \begin{split}
  f({\rm log}_{10}\,a) = \sqrt{\frac{2}{\pi}}\frac{({\rm log}_{10}\,a -  {\rm log}_{10}\,a_{\rm min})^2}{b^3}\\
  {\rm exp}\left[{\frac{-({\rm log}_{10}\,a - {\rm log}_{10}\,a_{\rm min})^2}{2b^2}}\right],
  \end{split} 
\end{equation}
where ${\rm log}_{10}\,a$ is the separation ($a \geq a_{\rm min}$), and $b$ is the characteristic width of the distribution\footnote{This is usually denoted $a$, but we avoid confusion with the semimajor axis by using $b$.}.

We keep $b$ constant, adopting $b = 2$, but we vary the peak of the distribution for each set of binaries by setting a different minimum separation  ${\rm log}_{10}\,a_{\rm min}$, as outlined in upper portion of Table~\ref{MB_fits_table}, with each set labelled `Constituent 1' -- `Constituent 5'. This allows the distribution to be moved along the ${\rm log}_{10}\,a$ axis such that it can have a different peak separation, but retain the shape of a Maxwell-Boltzmann distribution. 

We assume that each set of binaries is part of a population of stars with an overall multiplicity fraction $f_{\rm mult} = 0.55$, and separations drawn from the Maxwell-Boltzmann--like distributions summarised in Table~\ref{MB_fits_table}.

\begin{table}
  \caption{Different Maxwell-Boltzmann--like distribution fits used in this work.  We show the fit label (for interpretation of Figs.~\ref{Comparison_data}~and~\ref{MB_fits}), the width the the Maxwell-Boltzmann distribution $b$, the minimum (${\rm log}_{10}$) separation, ${\rm log}_{10}\,a_{\rm min}$, of the distribution, the resultant mean separation $\bar{a}$, and the logarithm of the mean (${\rm log}_{10}\,\bar{a}$), which is the peak of each Maxwell-Boltzmann distribution, and the multiplicity fraction $f_{\rm mult}$ the distribution is scaled to.}
\begin{center}
\begin{tabular}{cccccc}
\hline 

Label & $b$ & ${\rm log}_{10}\,a_{\rm min}$  & $\bar{a}$ & ${\rm log}_{10}\,\bar{a}$ & $f_{\rm mult}$ \\
\hline
Constituent 1 & 2 & $-3$ & 1.14\,au & 0.057  & 0.55\\
Constituent 2 & 2 & $-2.4$ & 3.72 & 0.57 &  0.55\\
Constituent 3 & 2 & $-1.8$ & 37\,au & 1.57 &  0.55\\
Constituent 4 & 2 & $-0.6$ & 372\,au & 2.57 &  0.55\\
Constituent 5 & 2 & $0.6$ & 3715\,au & 3.57 &  0.55 \\

\hline
MB fit 1 & 1.75 & $-2$ & 3.2\,au & 0.57 &  0.9\\
MB fit 2 & 1.25 & $-1$ & 6.3\,au & 0.80  & 0.75\\
MB fit 3 & 1.75 & $-1.5$ & 10\,au & 1.00 & 0.55 \\
\hline

\end{tabular}
\end{center}
\label{MB_fits_table}
\end{table}

\begin{figure}
 	\includegraphics[width=\columnwidth]{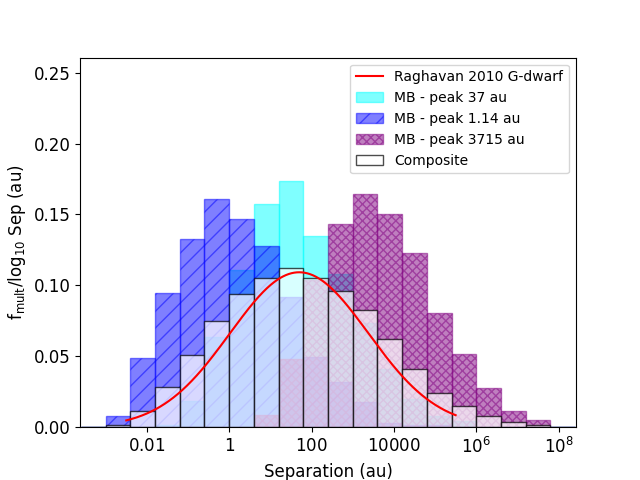}
        \caption{The separation distribution as a result of summing together five separate binary populations drawn from Maxwell-Boltzmann distributions with the properties listed in Table~\ref{MB_fits_table}. For clarity, we only show three of the Maxwell-Boltzmann distributions (those with peaks at 1, 37 and 3715\,au), but the summed distribution (the white/open histogram) contains all five populations, including those with peaks at 3.7 and 372\,au. For reference, we also show the log-normal fit to the field binaries presented in \citet{Raghavan10} by the solid red line. }
     \label{Comparison_data}
 \end{figure}

We show the summed distribution which combines all five binary populations by the open/white histogram in Fig.~\ref{Comparison_data}, and we show three of the five Maxwell-Boltzmann distributions (Constituent 1, Constituent 3 and Constituent 5 -- we omit the other two for clarity, but the binaries drawn from them are included in the summed distribution (the open/white histogram)). For comparison with the binary population in the Galactic field, we show the log-normal fit to the field from \citet{Raghavan10} by the solid red line.  Clearly, the mixture of these constituent Maxwell-Boltzmann distributions can produce a log-normal distribution similar to that in the field.

This demonstration of the central limit theorem  -- where the sum of several distributions leads to a log-normal \citep[e.g.][]{Hall80} -- suggests that the field separation distribution is the sum of many different separation distributions. Whilst we could make our composite population more complex by adding further constituent populations, and/or changing the multiplicity fraction of each population, we consider the summed composite distribution in Fig.~\ref{Comparison_data} an adequate demonstration of our argument.

We note here that our summed distribution involves contributions from constituent distributions with very different properties (i.e. much larger separations) than are predicted by theory \citep[e.g.][]{Bate12,Bate14} or observed in star-forming regions \citep[e.g.][]{Kohler98,Reipurth07,Duchene18}. Our goal here was to demonstrate that constituent distributions that deviate from a log-normal can reproduce a log-normal distribution as observed in the Galactic field. In reality, the widest binaries in the field are likely to form during the dynamical dissolution of star-forming regions \citep[e.g.][]{Kouwenhoven10,Moeckel10,Parker14d}, and there are very few constraints on the amount of possible variation in the separation distributions of close ($<$1\,au) binaries \citep{Stahler10,Korntreff12}.

 Having demonstrated that summed/combined Maxwell-Boltzmann distributions could result in a field-like log-normal distribution, we now show the type of fit that would be required to fit a smooth, continuous function to the data for the ONC, which now includes the 10\,au - 62\,au bin with a significant excess of binaries compared to both the Galactic field, and the wider binaries in the ONC with separations in the range 62--620\,au.

\begin{figure}
 	\includegraphics[width=\columnwidth]{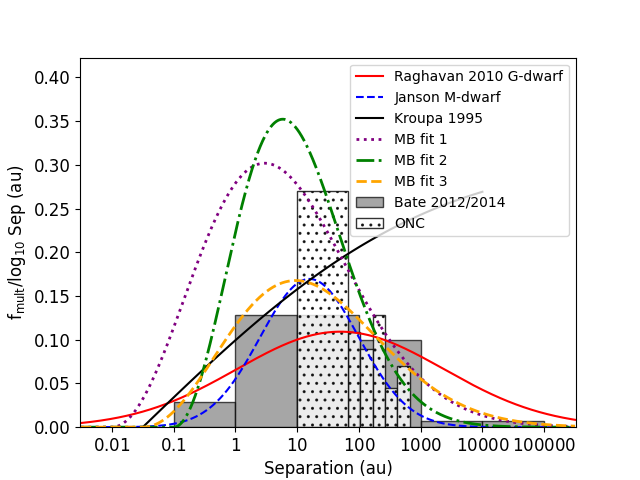}
        \caption{Hypothetical fits to the ONC binary separation distribution that would be consistent with the corresponding distributions from numerical simulations. The data for binaries in the ONC are shown by the dot-filled histogram. For comparison, we also show the log-normal fits to the field G-dwarf binaries \citep[][the solid red line]{Raghavan10} and the field M-dwarf binaries \citep[][the dashed blue line]{Bergfors10,Janson12}. The postulated `Universal' pre-main sequence from \citet{Kroupa95a} is shown by the solid black line. We show the simulation data from \citet{Bate12,Bate14} by the solid grey histogram. The two different Maxwell-Boltzmann distributions that would be consistent with the ONC data  are shown by the dotted purple line and the dot-dashed green line, and a Maxwell-Boltzmann distribution that is a reasonable fit to the separation distributions from numerical simulations is shown by the orange dashed line. Details for all three fits are provided in Table~\ref{MB_fits_table}. }
     \label{MB_fits}
 \end{figure}

 In Fig.~\ref{MB_fits} we show the data from the ONC by the dotted histogram, as well as the the log-normal fits to the field G-dwarf binaries \citep[][the solid red line]{Raghavan10} and the field M-dwarf binaries \citep[][the dashed blue line]{Bergfors10,Janson12}. The postulated `universal' pre-main sequence from \citet{Kroupa95a} is shown by the solid black line, and the grey histogram shows the separation distribution from hydrodynamical simulations \citep{Bate12,Bate14}.

 We then also show three Maxwell-Boltzmann--like distributions, two of which could feasibly fit the ONC data and and one that is more consistent with the results from numerical simulations \citep{Bate12,Bate14}. Details of the fits are provided in the lower portion of Table~\ref{MB_fits_table}. The first fit -- shown by the purple dot-dashed line in Fig.~\ref{MB_fits} -- has a peak at 3.2\,au and is normalised to a multiplicity fraction of $f_{\rm mult} = 0.9$. The second fit -- shown by the green dot-dashed line -- has a peak at 6.3\,au and is normalised to a binary fraction of $f_{\rm mult} = 0.75$. The third fit -- shown by the orange dashed line -- has a peak at 10\,au and is normalised to a binary fraction of 0.55.

 Both our fits to the ONC data peak at separations less than 10\,au (for context, the fits to the Galactic  field population peak at several tens of au), and the multiplicity fraction is significantly higher than that in the Galactic field (0.75--0.9 versus 0.3--0.6, depending on the study). The Maxwell-Boltzmann fit peaks at slightly shorter separations than the ONC data in the range 10--62\,au from \citet{Duchene18}, where there is a significant excess of binaries compared to the field. The tail of the Maxwell-Boltzmann fit then passes through the histogram bins for the wider systems in the ONC \citep[62 -- 620\,au,][]{Reipurth07}.

 In order to be consistent with the fits to the ONC data, the number of binaries in numerical simulations with separations $<50$\,au would need to be higher. The numbers of binaries in these simulations \emph{may} be a lower limit to true number because the simulation resolution limit is at around 0.5\,au (meaning more close binaries would form in reality). However, the initial conditions of the simulations will almost certainly be (slightly) different to the initial conditions for star formation in the ONC, which will affect the numbers of binaries that form, even if only at the level of statistical noise.

\section{Discussion}
\label{sec:discussion}

We have shown that a combination of binary populations where the separations are drawn from Maxwell-Boltzmann distributions can sum together to produce the log-normal distribution observed in the Galactic field binary population. We also show that a Maxwell-Boltzmann distribution could fit the observed separation distribution of binaries in the ONC, if the excess of binaries in the range 10--62\,au observed by \citet{Duchene18} continues to smaller separations.

We emphasise here that we are not proposing that a Maxwell-Boltzmann distribution -- which formally describes the velocities of particles in an ideal gas -- is a wholly appropriate distribution to decribe the semimajor axis distribution of binary stars \citep[see also][]{Jeans19,Ambartsumian37,Makarov25}. Rather, we propose that the characteristic shape of this distribution -- a peak with a tail of higher values -- is a better approximation to some observational and simulation data than the more commonly used (symmetrical) log-normal distribution. 

The implications of our results are two-fold. First, the excess of binaries in the range 10 -- 62\,au in the ONC does not necessarily imply that star-forming regions like the ONC do not contribute binary and multiple systems to the field. It is possible that further dynamical interactions in the ONC could reduce the number of these systems before they enter the field. However, this would also further reduce the number of wider (softer) binaries, thus making the overall distribution very un-field-like. If no (or very little) further dynamical processing of the ONC binary population occurs, then to be consistent with the field population, the ONC binaries would need to be combined with a population that had/has a deficit of system in the separation range 10 -- 62\,au.

The second implication of our results in some ways solves this first issue. A combination of binary populations with different peaks in their separation distributions (and potentially different multiplicity fractions) can result in a summed distribution with a very different shape, and peak separation, to some of the constituent distributions (e.g. as we show in Fig.~\ref{Comparison_data}). In the case of the ONC, it, or a star-forming region with a similar binary population, could still contribute to the field population if other contributing region(s) had a deficit of binaries in a similar separation range.

The separation distribution in the field spans seven orders of magnitude, from $10^{-2}$\,au to $10^5$\,au, whereas observations of binary populations in star-forming regions typically probe a much narrower range \citep[typically 10 -- 1000\,au, e.g.][]{Kohler98,Patience02,Ratzka05,Reipurth07,Kohler08,Duchene18}. That we can hide very un-field-like separation distributions within a field-like distribution, and that the ONC data can be fit with a Maxwell-Boltzmann distribution that would imply a significant excess of binaries with separations $<10$\,au, suggests that there are very few constraints for predicting the binary populations in individual star-forming regions.

The only targetted study of spectroscopic binaries in young star-forming regions is the work by \citet{Kounkel19}, who find little evidence of any variation in the multiplicity fraction compared to the field at small separations, and they suggest that the period (separation) distribution is also consistent with the field. However, this appears to be incongruous with the observed excess of 10 -- 62\,au separation binaries in the ONC \citep{Duchene18}, and in other star-forming regions \citep[see][for a review]{Offner23}. Furthermore, observations of spectroscopic binaries in nearby young moving groups suggest an excess of close binaries with respect to the field population \citep{Zuniga21}.

These results suggest that new observations targetted at the separation ranges currently not probed by observations of visual binaries in star-forming regions would be highly beneficial. The currently observed separation  range of visual binaries in star-forming regions is too narrow to draw any strong conclusions about the provenance of binary systems in the Galactic field, nor is it wide enough to robustly test the outcome of hydrodynamic simulations of star formation.


\section{Conclusions}
\label{sec:conclusions}

In this paper we have sampled asymmetric Maxwell-Boltzmann distributions to create a population of binary stars whose separation distribution resembles the log-normal separation distribution observed in the Galactic field. We compare the constituent distributions to the separation distribution observed in the ONC, and to the separation distribution produced in numerical simulations of star formation. Our conclusions are as follows:

(i) Binary populations with separations drawn from several Maxwell-Boltzmann distributions with different peak separations can be combined to form a log-normal distribution that resembles the Galactic field binary population. This implies that the field may not simply be the sum of lots of constituent distributions with similar properties (shape, mean separation, etc.), but that the constiuent distributions could have very different properties.

(ii) Similarly, the proposed `universal' separation distribution \citep{Kroupa95a,Kroupa11}, which posits that all binaries form from the same underlying distribution, requires dynamical processing of wide binaries to be important, and therefore requires most star-forming regions to be dense. If the field can be made of binaries drawn from different populations (i.e.\,\,in terms of the shape of the separation distribution and multiplicity fraction), then there are few -- if any -- constraints on the density of the the regions they formed in.

(iii) The observed excess of 10 -- 62\,au binaries in the ONC compared to both the Galactic field, and other star-forming regions, suggests that either the ONC would not contribute binaries to the field population, or that binary populations in star-forming regions are so different from one another that these differences are statistically insignificant in the sense that they are then lost in the overall combined distribution.

(iv) The binary separation distribution in the ONC can be fit with a Maxwell-Boltzmann distribution (or a distribution like it), which would require a similar or even larger excess of binaries with separations less than 10\,au as the excess of 10 -- 62\,au binaries compared to the field population.

(v) Whilst observations that probe close binaries in star-forming regions are currently in disagreement over whether the separation distribution for these system is field-like, there are hints of an excess of these systems in the separation distribution of binaries formed in numerical simulations of star forming regions.

Taken together, it is clear that the current observed range of binary separations in star-forming regions (typically 10s -- 1000\,au, compared to the $10^{-2} - 10^5$\,au range in the Galactic  field) is insufficient to constrain theories of multiple star formation, and further observations of (close) spectroscopic binaries ($a < 10$\,au) similar to the work by \citet{Kounkel19} and \citet{Zuniga21}, are urgently needed.

\section*{Acknowledgements}

We thank the anonymous referee for their helpful and constructive reports.  MHR was funded by a 2024 Sheffield Undergraduate Research Experience (SURE) bursary. RJP acknowledges support from the Royal Society in the form of a Dorothy Hodgkin research fellowship. For the purpose of open access, the authors have applied a Creative Commons Attribution (CC BY) licence to any Author Accepted Manuscript version arisin

\section*{Data Availability}

No new data were generated in this work. 



\bibliographystyle{aasjournal}
\bibliography{general_ref} 

\begin{thebibliography}{}
\expandafter\ifx\csname natexlab\endcsname\relax\def\natexlab#1{#1}\fi
\providecommand{\url}[1]{\href{#1}{#1}}
\providecommand{\dodoi}[1]{doi:~\href{http://doi.org/#1}{\nolinkurl{#1}}}
\providecommand{\doeprint}[1]{\href{http://ascl.net/#1}{\nolinkurl{http://ascl.net/#1}}}
\providecommand{\doarXiv}[1]{\href{https://arxiv.org/abs/#1}{\nolinkurl{https://arxiv.org/abs/#1}}}

\bibitem[{{Ambartsumian}(1937)}]{Ambartsumian37}
{Ambartsumian}, V.~A. 1937, \azh, 14, 207

\bibitem[{{Ballantyne} {et~al.}(2021){Ballantyne}, {Espaas}, {Norgrove},
  {Wootton}, {Harris}, {Pepper}, {Smith}, {Dommett}, \&
  {Parker}}]{Ballantyne21}
{Ballantyne}, H.~A., {Espaas}, T., {Norgrove}, B.~Z., {et~al.} 2021, \mnras,
  507, 4507, \dodoi{10.1093/mnras/stab2324}

\bibitem[{Bate(2012)}]{Bate12}
Bate, M.~R. 2012, MNRAS, 419, 3115

\bibitem[{{Bate}(2014)}]{Bate14}
{Bate}, M.~R. 2014, \mnras, 442, 285, \dodoi{10.1093/mnras/stu795}

\bibitem[{{Bergfors} {et~al.}(2010){Bergfors}, Brandner, Janson, Daemgen,
  Geissler, Henning, Hippler, Hormuth, Joergens, \& K{\"o}hler}]{Bergfors10}
{Bergfors}, C., Brandner, W., Janson, M., {et~al.} 2010, A\&A, 520, A54

\bibitem[{{Chen} {et~al.}(2013){Chen}, {Arce}, {Zhang}, {Bourke}, {Launhardt},
  {J{\o}rgensen}, {Lee}, {Foster}, {Dunham}, {Pineda}, \& {Henning}}]{Chen13}
{Chen}, X., {Arce}, H.~G., {Zhang}, Q., {et~al.} 2013, ApJ, 768, 110,
  \dodoi{10.1088/0004-637X/768/2/110}

\bibitem[{{Cifuentes} {et~al.}(2024){Cifuentes}, {Caballero},
  {Gonz{\'a}lez-Payo}, {Amado}, {B{\'e}jar}, {Burgasser},
  {Cort{\'e}s-Contreras}, {Lodieu}, {Montes}, {Quirrenbach}, {Reiners},
  {Ribas}, {Sanz-Forcada}, {Seifert}, \& {Zapatero Osorio}}]{Cifuentes25}
{Cifuentes}, C., {Caballero}, J.~A., {Gonz{\'a}lez-Payo}, J., {et~al.} 2024,
  arXiv e-prints, arXiv:2412.12264, \dodoi{10.48550/arXiv.2412.12264}

\bibitem[{{Cournoyer-Cloutier} {et~al.}(2024){Cournoyer-Cloutier}, {Sills},
  {Harris}, {Polak}, {Rieder}, {Andersson}, {Appel}, {Mac Low}, {McMillan}, \&
  {Portegies Zwart}}]{Cloutier24}
{Cournoyer-Cloutier}, C., {Sills}, A., {Harris}, W.~E., {et~al.} 2024, \apj,
  977, 203, \dodoi{10.3847/1538-4357/ad90b3}

\bibitem[{{De Rosa} {et~al.}(2014){De Rosa}, Patience, Vigan, Wilson,
  Schneider, McConnell, Wiktorowicz, Marois, Song, Macintosh, Graham, Bessell,
  Doyon, Lai, \& Thomas}]{DeRosa14}
{De Rosa}, R.~J., Patience, J., Vigan, A., {et~al.} 2014, MNRAS, 437, 1216

\bibitem[{{Duch{\^e}ne} \& {Kraus}(2013)}]{Duchene13b}
{Duch{\^e}ne}, G., \& {Kraus}, A. 2013, ARA\&A, 51, 269,
  \dodoi{10.1146/annurev-astro-081710-102602}

\bibitem[{{Duch{\^e}ne} {et~al.}(2018){Duch{\^e}ne}, {Lacour}, {Moraux},
  {Goodwin}, \& {Bouvier}}]{Duchene18}
{Duch{\^e}ne}, G., {Lacour}, S., {Moraux}, E., {Goodwin}, S., \& {Bouvier}, J.
  2018, \mnras, 478, 1825, \dodoi{10.1093/mnras/sty1180}

\bibitem[{Duquennoy \& Mayor(1991)}]{Duquennoy91}
Duquennoy, A., \& Mayor, M. 1991, A\&A, 248, 485

\bibitem[{{El-Badry} {et~al.}(2021){El-Badry}, {Rix}, \& {Heintz}}]{ElBadry21}
{El-Badry}, K., {Rix}, H.-W., \& {Heintz}, T.~M. 2021, \mnras, 506, 2269,
  \dodoi{10.1093/mnras/stab323}

\bibitem[{Fischer \& Marcy(1992)}]{Fischer92}
Fischer, D.~A., \& Marcy, G.~W. 1992, ApJ, 396, 178

\bibitem[{Hall \& Heyde(1980)}]{Hall80}
Hall, P., \& Heyde, C. 1980, in Martingale Limit Theory and its Application,
  ed. P.~Hall \& C.~Heyde, Probability and Mathematical Statistics: A Series of
  Monographs and Textbooks (Academic Press), 51--96,
  \dodoi{https://doi.org/10.1016/B978-0-12-319350-6.50009-8}

\bibitem[{Heggie(1975)}]{Heggie75}
Heggie, D.~C. 1975, MNRAS, 173, 729

\bibitem[{Hills(1975{\natexlab{a}})}]{Hills75a}
Hills, J.~G. 1975{\natexlab{a}}, AJ, 80, 809

\bibitem[{Hills(1975{\natexlab{b}})}]{Hills75b}
---. 1975{\natexlab{b}}, AJ, 80, 1075

\bibitem[{Janson {et~al.}(2012)Janson, Hormuth, Bergfors, Brandner, Hippler,
  Daemgen, Kudryavtseva, Schmalzl, Schnupp, \& Henning}]{Janson12}
Janson, M., Hormuth, F., Bergfors, C., {et~al.} 2012, ApJ, 754, 44

\bibitem[{Jeans(1919)}]{Jeans19}
Jeans, J.~H. 1919, MNRAS, 79, 408

\bibitem[{King {et~al.}(2012{\natexlab{a}})King, Goodwin, Parker, \&
  Patience}]{King12b}
King, R.~R., Goodwin, S.~P., Parker, R.~J., \& Patience, J. 2012{\natexlab{a}},
  MNRAS, 427, 2636

\bibitem[{King {et~al.}(2012{\natexlab{b}})King, Parker, Patience, \&
  Goodwin}]{King12a}
King, R.~R., Parker, R.~J., Patience, J., \& Goodwin, S.~P. 2012{\natexlab{b}},
  MNRAS, 421, 2025

\bibitem[{K{\"o}hler \& Leinert(1998)}]{Kohler98}
K{\"o}hler, R., \& Leinert, C. 1998, A\&A, 331, 977

\bibitem[{K{\"o}hler {et~al.}(2008)K{\"o}hler, Neuh{\"a}user, Kr{\"a}mer,
  Leinert, Ott, \& Eckart}]{Kohler08}
K{\"o}hler, R., Neuh{\"a}user, R., Kr{\"a}mer, S., {et~al.} 2008, A\&A, 488,
  997

\bibitem[{{Korntreff} {et~al.}(2012){Korntreff}, {Kaczmarek}, \&
  {Pfalzner}}]{Korntreff12}
{Korntreff}, C., {Kaczmarek}, T., \& {Pfalzner}, S. 2012, A\&A, 543, A126,
  \dodoi{10.1051/0004-6361/201118019}

\bibitem[{{Kounkel} {et~al.}(2019){Kounkel}, {Covey}, {Moe}, {Kratter},
  {Su{\'a}rez}, {Stassun}, {Rom{\'a}n-Z{\'u}{\~n}iga}, {Hernandez}, {Kim},
  {Pe{\~n}a Ram{\'\i}rez}, {Roman-Lopes}, {Stringfellow}, {Jaehnig},
  {Borissova}, {Tofflemire}, {Krolikowski}, {Rizzuto}, {Kraus}, {Badenes},
  {Longa-Pe{\~n}a}, {G{\'o}mez Maqueo Chew}, {Barba}, {Nidever}, {Brown}, {De
  Lee}, {Pan}, {Bizyaev}, {Oravetz}, \& {Oravetz}}]{Kounkel19}
{Kounkel}, M., {Covey}, K., {Moe}, M., {et~al.} 2019, \aj, 157, 196,
  \dodoi{10.3847/1538-3881/ab13b1}

\bibitem[{Kouwenhoven {et~al.}(2010)Kouwenhoven, Goodwin, Parker, Davies,
  Malmberg, \& Kroupa}]{Kouwenhoven10}
Kouwenhoven, M. B.~N., Goodwin, S.~P., Parker, R.~J., {et~al.} 2010, MNRAS,
  404, 1835

\bibitem[{Kroupa(1995{\natexlab{a}})}]{Kroupa95a}
Kroupa, P. 1995{\natexlab{a}}, MNRAS, 277, 1491

\bibitem[{Kroupa(1995{\natexlab{b}})}]{Kroupa95b}
---. 1995{\natexlab{b}}, MNRAS, 277, 1507

\bibitem[{Kroupa {et~al.}(1999)Kroupa, Petr, \& McCaughrean}]{Kroupa99}
Kroupa, P., Petr, M.~G., \& McCaughrean, M.~J. 1999, New Astronomy, 4, 495

\bibitem[{Kroupa \& {Petr-Gotzens}(2011)}]{Kroupa11}
Kroupa, P., \& {Petr-Gotzens}, M.~G. 2011, A\&A, 529, A92

\bibitem[{{Makarov}(2025)}]{Makarov25}
{Makarov}, V.~V. 2025, arXiv e-prints, arXiv:2501.02587,
  \dodoi{10.48550/arXiv.2501.02587}

\bibitem[{{Marks} \& {Kroupa}(2012)}]{Marks12}
{Marks}, M., \& {Kroupa}, P. 2012, A\&A, 543, A8,
  \dodoi{10.1051/0004-6361/201118231}

\bibitem[{Marks {et~al.}(2011)Marks, Kroupa, \& Oh}]{Marks11}
Marks, M., Kroupa, P., \& Oh, S. 2011, MNRAS, 417, 1684

\bibitem[{Moeckel \& Bate(2010)}]{Moeckel10}
Moeckel, N., \& Bate, M.~R. 2010, MNRAS, 404, 721

\bibitem[{{Offner} {et~al.}(2023){Offner}, {Moe}, {Kratter}, {Sadavoy},
  {Jensen}, \& {Tobin}}]{Offner23}
{Offner}, S.~S.~R., {Moe}, M., {Kratter}, K.~M., {et~al.} 2023, in Astronomical
  Society of the Pacific Conference Series, Vol. 534, Protostars and Planets
  VII, ed. S.~{Inutsuka}, Y.~{Aikawa}, T.~{Muto}, K.~{Tomida}, \& M.~{Tamura},
  275, \dodoi{10.48550/arXiv.2203.10066}

\bibitem[{{Parker}(2023)}]{Parker23d}
{Parker}, R.~J. 2023, \mnras, 525, 2907, \dodoi{10.1093/mnras/stad2444}

\bibitem[{Parker {et~al.}(2011)Parker, Goodwin, \& Allison}]{Parker11c}
Parker, R.~J., Goodwin, S.~P., \& Allison, R.~J. 2011, MNRAS, 418, 2565

\bibitem[{Parker \& Meyer(2014)}]{Parker14d}
Parker, R.~J., \& Meyer, M.~R. 2014, MNRAS, 442, 3722

\bibitem[{Patience {et~al.}(2002)Patience, Ghez, Reid, \&
  Matthews}]{Patience02}
Patience, J., Ghez, A.~M., Reid, I.~N., \& Matthews, K. 2002, AJ, 123, 1570

\bibitem[{Raghavan {et~al.}(2010)Raghavan, McMaster, Henry, Latham, Marcy,
  Mason, Gies, White, \& {ten Brummelaar}}]{Raghavan10}
Raghavan, D., McMaster, H.~A., Henry, T.~J., {et~al.} 2010, ApJSS, 190, 1

\bibitem[{{Ratzka} {et~al.}(2005){Ratzka}, {K{\"o}hler}, \&
  {Leinert}}]{Ratzka05}
{Ratzka}, T., {K{\"o}hler}, R., \& {Leinert}, C. 2005, \aap, 437, 611,
  \dodoi{10.1051/0004-6361:20042107}

\bibitem[{Reipurth {et~al.}(2007)Reipurth, Guimar{\~a}es, Connelley, \&
  Bally}]{Reipurth07}
Reipurth, B., Guimar{\~a}es, M.~M., Connelley, M.~S., \& Bally, J. 2007, AJ,
  134, 2272

\bibitem[{{Sana} {et~al.}(2013){Sana}, {de Koter}, {de Mink}, {Dunstall},
  {Evans}, \& {et al.}}]{Sana13}
{Sana}, H., {de Koter}, A., {de Mink}, S.~E., {et~al.} 2013, A\&A, 550, A107,
  \dodoi{10.1051/0004-6361/201219621}

\bibitem[{{Stahler}(2010)}]{Stahler10}
{Stahler}, S.~W. 2010, MNRAS, 402, 1758,
  \dodoi{10.1111/j.1365-2966.2009.15994.x}

\bibitem[{{Tokovinin}(2008)}]{Tokovinin08}
{Tokovinin}, A. 2008, MNRAS, 389, 925, \dodoi{10.1111/j.1365-2966.2008.13613.x}

\bibitem[{{Tokovinin}(2014)}]{Tokovinin14}
---. 2014, AJ, 147, 87, \dodoi{10.1088/0004-6256/147/4/87}

\bibitem[{{Ward-Duong} {et~al.}(2015){Ward-Duong}, {Patience}, {De Rosa},
  {Bulger}, {Rajan}, {Goodwin}, {Parker}, {McCarthy}, \&
  {Kulesa}}]{Ward-Duong15}
{Ward-Duong}, K., {Patience}, J., {De Rosa}, R.~J., {et~al.} 2015, MNRAS, 449,
  2618.
\newblock \doarXiv{1503.00724}

\bibitem[{{Z{\'u}{\~n}iga-Fern{\'a}ndez}
  {et~al.}(2021){Z{\'u}{\~n}iga-Fern{\'a}ndez}, {Bayo}, {Elliott}, {Zamora},
  {Corval{\'a}n}, {Haubois}, {Corral-Santana}, {Olofsson}, {Hu{\'e}lamo},
  {Sterzik}, {Torres}, {Quast}, \& {Melo}}]{Zuniga21}
{Z{\'u}{\~n}iga-Fern{\'a}ndez}, S., {Bayo}, A., {Elliott}, P., {et~al.} 2021,
  \aap, 645, A30, \dodoi{10.1051/0004-6361/202037830}

\end{thebibliography}








\end{document}